\documentclass[
 aip,
 pop,
amsfonts,
amsmath,
amssymb,
reprint,%
]{revtex4-1}
\usepackage{subfig}
\usepackage{float}
\usepackage{multirow}
\usepackage{graphicx}
\usepackage{dcolumn}
\usepackage{bm}
\usepackage{color}
\usepackage[colorlinks,linkcolor=blue,urlcolor=blue,citecolor=blue]{hyperref}
\usepackage[table]{xcolor}

\begin{document}
\title{Accuracy of Kohn-Sham density functional theory for warm- and hot-dense matter equation of state}

\author{Phanish Suryanarayana}
\email[]{phanish.suryanarayana@ce.gatech.edu}
\affiliation{College of Engineering, Georgia Institute of Technology, Atlanta, Georgia 30332, USA}
\affiliation{College of Computing, Georgia Institute of Technology, Atlanta, Georgia 30332, USA}

\author{Arpit Bhardwaj}
\affiliation{College of Engineering, Georgia Institute of Technology, Atlanta, Georgia 30332, USA}

\author{Xin Jing}
\affiliation{College of Engineering, Georgia Institute of Technology, Atlanta, Georgia 30332, USA}
\affiliation{College of Computing, Georgia Institute of Technology, Atlanta, Georgia 30332, USA}

\author{Shashikant Kumar}
\affiliation{College of Engineering, Georgia Institute of Technology, Atlanta, Georgia 30332, USA}

\author{John E. Pask}
\affiliation{Physics Division, Lawrence Livermore National Laboratory, Livermore, California 94550, USA}


\date{\today}

\begin{abstract}
We study the accuracy of Kohn-Sham density functional theory (DFT) for warm- and hot-dense matter (WDM and HDM). Specifically,  considering a wide range of systems, we perform accurate ab initio molecular dynamics simulations with temperature-independent local/semilocal density functionals to determine the equations of state at compression ratios of 3x--7x and temperatures near 1 MK. We find very good agreement with path integral Monte Carlo benchmarks, while having significantly smaller error bars and smoother data,  demonstrating the accuracy of DFT for the study of WDM and HDM at such conditions. In addition, using a $\Delta$-machine learned force field scheme, we confirm that the DFT results are insensitive to the choice of exchange-correlation functional, whether local, semilocal, or nonlocal.
\end{abstract}

\maketitle



\section{Introduction}

Warm- and hot-dense matter (WDM and HDM) arise in diverse physical systems such as  the interiors of stars and giant planets as well as inertial confinement fusion (ICF) and other high energy density (HED) experiments \cite{gradesred2014}. The ability to accurately model WDM and HDM is therefore a key component in the understanding and design of ICF experiments as well as understanding the formation, nature, and evolution of planetary/stellar systems. However, such extreme conditions of temperature and pressure pose a significant  challenge experimentally as well as theoretically since both classical and quantum mechanical (degeneracy) effects occur, with the relative dominance of each varying with density and temperature.

Path integral Monte Carlo (PIMC) is a many-body  method for the study of materials systems from the first principles of quantum mechanics, without any empirical or adjustable parameters \cite{barker1979quantum}. Though highly accurate and capable of producing benchmark results, the PIMC method has significant computational cost. This cost, however, decreases with increase in temperature, allowing for  the study of systems at temperatures associated with  WDM/HDM \cite{driver2018path, zhang2018theoretical, zhang2020benchmarking, zhang2019equation, driver2012all, benedict2014multiphase, zhang2018path, zhang2017first, militzer2001path, hu2011first, militzer1999characterization, militzer2009path, militzer2006first, driver2017first, gonzalez2020equation, soubiran2019magnesium, gonzalez2020path, driver2016first, zhang2017equation, driver2015first, driver2015oxygen, militzer2015development, militzer2021first}. However, despite significant advances, PIMC calculations remain limited in practice to small systems of low-Z elements at temperatures at or above the upper (lower) end of WDM (HDM) conditions. Furthermore, quantities of interest such as ionic transport properties are not accessible. 

Kohn-Sham density functional theory (DFT) \cite{Kohn1965, Hohenberg, mermin1965thermal} is among the most widely used first principles methods for the study of materials systems.  Given its high accuracy-to-cost ratio relative to other ab initio methods such as PIMC, DFT has become a cornerstone of modern materials science research, particularly for systems at or near ambient conditions \cite{burke2012perspective}. However, Kohn-Sham calculations typically scale quadratically with the number of states computed for a fixed cell size \cite{martin2020electronic}, making calculations at high temperature particularly expensive given the increase in partially occupied states with temperature. This bottleneck is particularly acute in ab initio molecular dynamics (AIMD) simulations, where the Kohn-Sham equations may need to be solved thousands or hundreds of thousands of times to reach time scales of interest \cite{burke2012perspective}. Consequently, conventional diagonalization-based Kohn-Sham DFT has found rather limited use in the study of WDM and HDM \cite{bethkenhagen2023properties, wu2021development, driver2018path, zhang2018theoretical, zhang2020benchmarking, zhang2019equation, driver2012all, benedict2014multiphase, zhang2018path, zhang2017first, militzer2009path, militzer2006first, driver2017first, gonzalez2020equation, soubiran2019magnesium, gonzalez2020path, driver2016first, zhang2017equation, driver2015first, driver2015oxygen, militzer2015development, militzer2021first}. This has motivated the development of new solution strategies to increase the efficiency of Kohn-Sham calculations under such conditions \cite{pratapa2016spectral, suryanarayana2017sqdft, sharma2020real, xu2022real, zhang2016extended, blanchet2021extended, cytter2018stochastic, white2020fast, sadigh2023spectral}. 
However, despite significant advances, the accuracy of Kohn-Sham DFT for WDM and HDM has not been clearly and comprehensively assessed heretofore. 

In this work, we quantify the accuracy of Kohn-Sham DFT for WDM and HDM by carrying out a comprehensive  comparison of highly accurate Kohn-Sham results to available PIMC benchmarks \cite{militzer2021first}. In particular, we employ the linear scaling Spectral Quadrature (SQ) method \cite{suryanarayana2013spectral, pratapa2016spectral, suryanarayana2017sqdft} implemented in the open-source SPARC electronic structure code \cite{xu2021sparc, zhang2023version, ghosh2016sparc2} to perform AIMD simulations to determine the equations of state (EOS) at compression ratios of 3x--7x and temperatures near 1 MK for a wide range of systems in the FPEOS database \cite{militzer2021first}. The SQ method enables systematic convergence of full Kohn-Sham energies, forces, and stresses to high accuracy at WDM/HDM conditions \cite{pratapa2016spectral, suryanarayana2017sqdft, sharma2020real}, without homogeneous electron gas (HEG) approximations \cite{zhang2016extended, blanchet2021extended, sadigh2023spectral} or large prefactors to reduce statistical errors to levels targeted in practice \cite{cytter2018stochastic, white2020fast}. Finally, using a $\Delta$-MLFF scheme, we confirm that the DFT results are insensitive to the choice of exchange-correlation functional, whether local, semilocal, or nonlocal, the first such demonstration at the fully nonlocal level to our knowledge.

\section{Methods}
We now provide a brief description of the methods used in this work for the calculation of the EOS, namely the Spectral Quadrature and $\Delta$-machine learned force field methods.

\subsection{Spectral Quadrature}
Neglecting spin and Brillouin zone integration, the single-particle finite-temperature density matrix in Kohn-Sham DFT \cite{Hohenberg, Kohn1965, mermin1965thermal} can be written as
\begin{align}
\mathcal{D} = f(\mathcal{H}) = \left(1 + \exp \left( \frac{\mathcal{H} - \mu \mathcal{I}}{k_B T} \right) \right)^{-1} \,,
\end{align}
where $f$ is the Fermi-Dirac function, $\mathcal{H}$ is the Hamiltonian matrix, $\mathcal{I}$ is the identity matrix, $k_B$ is the Boltzmann constant, $T$ is the temperature, and $\mu$ is the chemical potential, obtained by enforcing the constraint on the total number of electrons. The calculation of the density matrix represents a nonlinear  problem since the Hamiltonian itself depends on the components of the density matrix. 
Once the self-consistent solution has been obtained, the ground state density matrix can be used to compute quantities of interest such as the energy (free and internal), atomic forces, and  pressure.

The finite-temperature density matrix  has exponential decay away from its diagonal  in a real-space representation \cite{goedecker1998decay, ismail1999locality, benzi2013decay, suryanarayana2017nearsightedness}, a consequence of the locality of electronic interactions, commonly referred to as ``nearsightedness'' in many-atom systems \cite{prodan2005nearsightedness}. This decay can be exploited in the development of reduced scaling solution methods for  the Kohn-Sham problem \cite{bowler2002recent, Goedecker}. One such linear scaling method that is particularly well suited to calculations at high temperature is Spectral Quadrature (SQ)~\cite{suryanarayana2013spectral}. In the SQ method, the decay in the density matrix is exploited to approximate quantities of interest as spatially localized bilinear forms or sums of bilinear forms, which are then approximated by quadrature rules. For instance, within a real-space finite-difference representation, the diagonal of the density matrix is the electron density, whose value at the $n$-th finite-difference node can be approximated as
\begin{align} \label{Eq:Dens:SQ}
e_n^{\rm T} \mathcal{D} e_n \approx \sum_{k=1}^{N_q} w_{n,k}  f(\lambda_{n,k}) \,,
\end{align}
where $e_n$ is the standard basis vector, $N_q$ is the quadrature order, $\lambda_{n,k}$ are the quadrature nodes,  and $w_{n,k}$ are the corresponding quadrature weights.  The weights can be formally written as
\begin{align}\label{Eq:weights:SQ}
w_{n,k}  = e_s^{\rm T} L_k(\mathcal{H}_{n}) e_s  \,,
\end{align}
where $L_k$ are the Lagrange polynomials, and $e_s$ is the standard basis vector associated with the $n$-th finite-difference node subsequent to forming the truncated Hamiltonian $\mathcal{H}_{n}$ for that node. In practice, the quadrature weights are not evaluated using Eq.~\ref{Eq:weights:SQ} directly since the computational cost would scale quadratically with quadrature order. Rather, alternate polynomial bases whose product with the vector $e_s$ can be efficiently evaluated using three-term recurrence relations are used, whereby the computational cost scales linearly with the quadrature order. In so doing, off-diagonal components of the truncated density matrix also become available, which can be used to efficiently calculate quantities such as atomic forces and pressure. 

The SQ method is applicable to metallic and insulating systems alike, with increasing efficiency at higher temperature  as the density matrix becomes more localized and the Fermi operator becomes smoother \cite{suryanarayana2017sqdft, suryanarayana2017nearsightedness, pratapa2016spectral}. With increasing  quadrature order and localization radius, the exact diagonalization limit can be obtained to desired accuracy,  with convergence to standard diagonalization-based planewave results readily obtained \cite{sharma2020real, suryanarayana2017sqdft, pratapa2016spectral}. The SQ method also provides results for the infinite-crystal  without recourse to Brillouin zone integration or large supercells \cite{Phanish2012, suryanarayana2017sqdft, pratapa2016spectral}, a technique referred to as the infinite-cell method. 

\subsection{\label{sec:delta_method}$\Delta$-machine learned force field}
In the $\Delta$-machine learned force field (MLFF) formalism \cite{bowman2022delta, bogojeski2020quantum, nandi2021delta, qu2021breaking, liu2022phase, verdi2023quantum, kumar2023kohn}, the differences in free energy, atomic forces, and pressure between two levels of theory are approximated using a machine-learned model. In particular, the difference in free energy  can be  written as \cite{kumar2023kohn}
\begin{equation} 
    \mathcal{E}_{\Delta} = \sum_{E=1}^{N_{E}} \sum_{i=1}^{N_A^E} \sum_{t=1}^{N_{T}^{E}} \tilde{w}_{t}^{E} k\left(\mathbf{x}_{i}^{E}, \tilde{\mathbf{x}}_t^{E} \right) \,,
    \label{Eq:EnergyDecompML}
\end{equation}
where the index $E$ denotes the chemical element, with $N_{E}$ being the total number of such species. In addition, $N_A^E$ is the number of atoms, $N_T^E$ is the number of local atomic descriptors, $\tilde{w}_{t}^{E}$ are the unknown model weights, and $k\left(\mathbf{x}_{i}^{E},\tilde{\mathbf{x}}_t^{E} \right)$ is the kernel function that measures the similarity between the descriptor vectors $\mathbf{x}_{i}^{E}$ and $\tilde{\mathbf{x}}_t^{E}$, the latter corresponding to those in the training dataset. We utilize the fourth-order polynomial kernel \cite{bartok2013representing, jinnouchi2019phase}  and the Smooth Overlap of Atomic Positions (SOAP) descriptors\cite{bartok2013representing, bartok2010gaussian}. The differences in atomic forces and pressure are obtained by taking the derivative of free energy difference in Eq.~\ref{Eq:EnergyDecompML} with respect to atomic positions and volumetric strains, respectively, the expressions for which can be found in the literature \cite{kumar2023kohn}. 


The model weights can be calculated using  Bayesian linear regression:
\begin{equation} 
    \bm{\tilde{w}} = \beta(\alpha \bm{\mathit{I}} + \beta \bm{\tilde{K}}^T\mathbf{\tilde{K}})^{-1}  \bm{\tilde{K}}^{\rm T} \bm{\tilde{y}} \,,
    \label{Eq:wtsML}
\end{equation}
where $\bm{\tilde{w}}$ is a vector containing the weights; $\alpha$ and $\beta$ are parameters determined by maximizing the evidence function \cite{bishop2006pattern}; $\bm{\tilde{y}}$ is a vector containing the normalized quantities of interest, i.e., differences in free energy, atomic forces, and pressure, shifted and scaled by their respective mean and standard deviation, respectively; and $\bm{\tilde{K}}$ is the covariance matrix corresponding to the training configurations. Once the weights have been computed, i.e., model has been trained, the differences in free energy, atomic forces, and pressure for a new atomic configuration, denoted by the vector $\bm{y}$, can then be predicted as:
\begin{equation} 
    \bm{y} = \bm{k} \bm{\tilde{w}} \,,
    \label{Eq:LinearSystemML}
\end{equation}
where $\bm{k}$ is the covariance matrix associated with the new atomic configuration. Since the above formalism provides the free energy difference and not the internal energy difference between the two levels of theory, a linear regression model between the free energy difference and internal energy difference is used to compute the latter \cite{kumar2024fly}.

 
\section{Results and discussion}
We now assess the accuracy of Kohn-Sham DFT for the study of WDM and HDM. To do so, we first consider the following systems at compression ratios of 3x--7x and temperatures near 1 MK:  hydrogen (H), helium (He), boron (B), carbon (C),  nitrogen (N), oxygen (O), neon (Ne), magnesium (Mg), lithium fluoride (LiF), boron nitride (BN), hydrocarbon (CH), boron carbide (B$_{4}$C), and magnesium silicate (MgSiO$_{3}$). These constitute a representative set of systems from the first-principles equation of state (FPEOS) database \cite{militzer2021first} for which PIMC data is available for comparison, providing a stringent test of DFT's accuracy.

We determine the pressure and internal energy at a given density and temperature by performing isokinetic ensemble (NVK) AIMD simulations with a Gaussian thermostat \cite{minary2003algorithms}. We consider 32-atom cells for all systems other than B$_{4}$C and MgSiO$_{3}$, for which we consider 40-atom cells consistent with stoichiometry. We employ the temperature-independent local density approximation (LDA) as the exchange-correlation functional \cite{Kohn1965, perdew1992accurate}.  We employ optimized norm conserving Vanderbilt (ONCV) pseudopotentials \cite{hamann2013optimized} with $1$, $2$, $5$, $6$, $7$, $8$, $10$, $10$, $3$, $9$, and $12$ electrons in valence for H, He, B, C, N, O, Ne, Mg, Li, F, and Si, respectively.  These pseudopotentials, most of which have  all electrons in valence, have been constructed for target accuracy at the conditions considered. 

We perform the DFT calculations using the SQ method implemented in the open-source SPARC electronic structure code \cite{xu2021sparc, zhang2023version, ghosh2016sparc2}. In particular, Gauss SQ is used for calculation of the electron density and energy \cite{suryanarayana2013spectral, pratapa2016spectral, suryanarayana2017sqdft}, whereas Clenshaw-Curtis SQ is used for calculation of the forces and the pressure \cite{pratapa2016spectral, suryanarayana2017sqdft, sharma2020real}. Numerical parameters in the DFT calculation, including the grid spacing, quadrature order, and localization radius are chosen to converge pressures to 0.5$\%$ or less (discretization error). A time step of 0.04 fs is used in AIMD simulations, where the initial $\sim$ 500 steps are used for equilibration and the following $\sim$ 8000 steps  are used for production, which reduces error bars on the pressure to 0.05\% or less (statistical error).

\begin{figure*}[htbp!] 
\centering

\subfloat[B]{\includegraphics[keepaspectratio=true,width=0.32\textwidth]{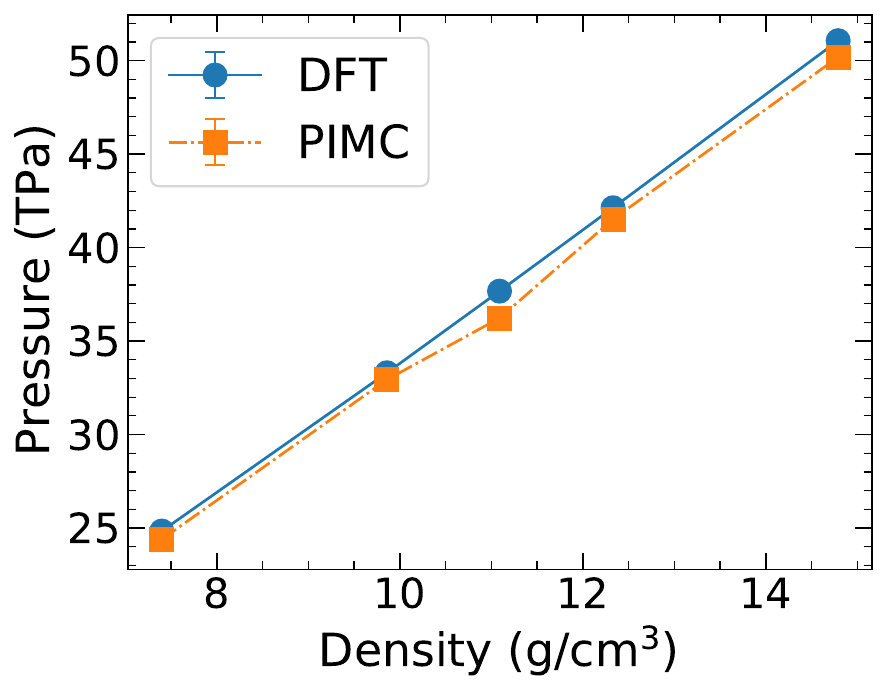}}\hspace{0.001\textwidth}
\subfloat[C]{\includegraphics[keepaspectratio=true,width=0.32\textwidth]{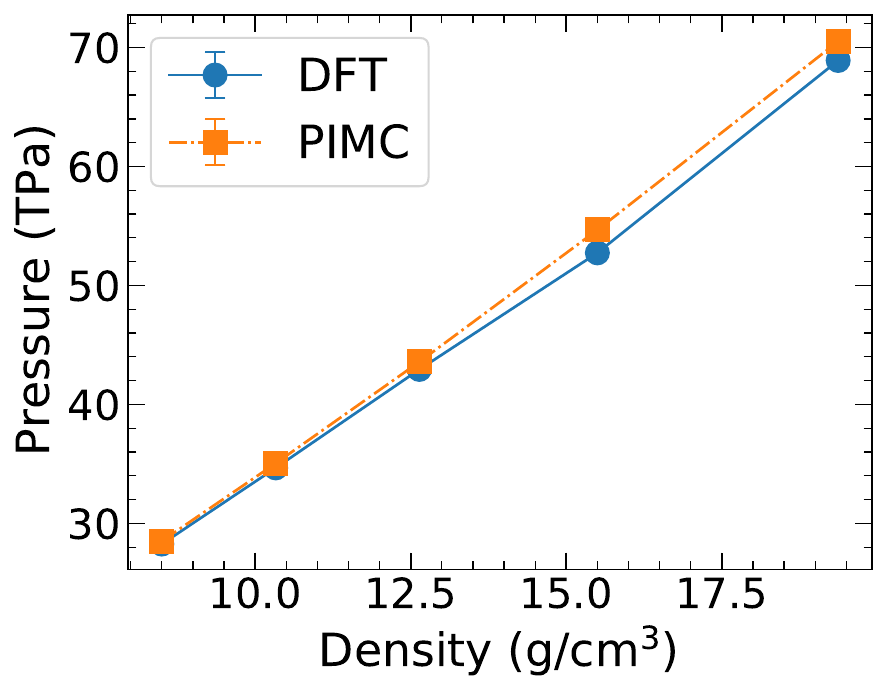}}\hspace{0.01\textwidth}
\subfloat[BN]{\includegraphics[keepaspectratio=true,width=0.32\textwidth]{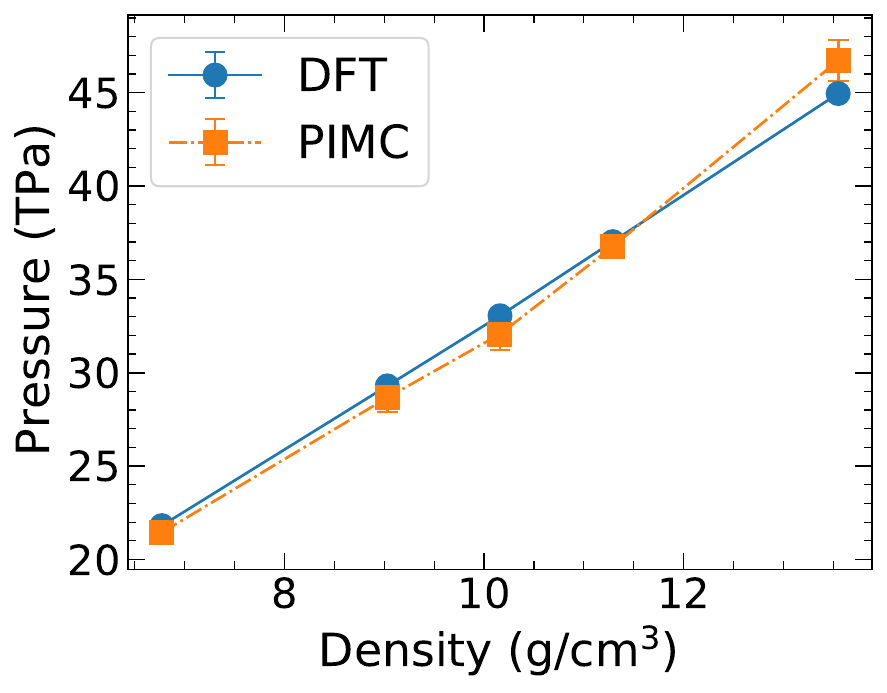}}

\subfloat[CH]{\includegraphics[keepaspectratio=true,width=0.32\textwidth]{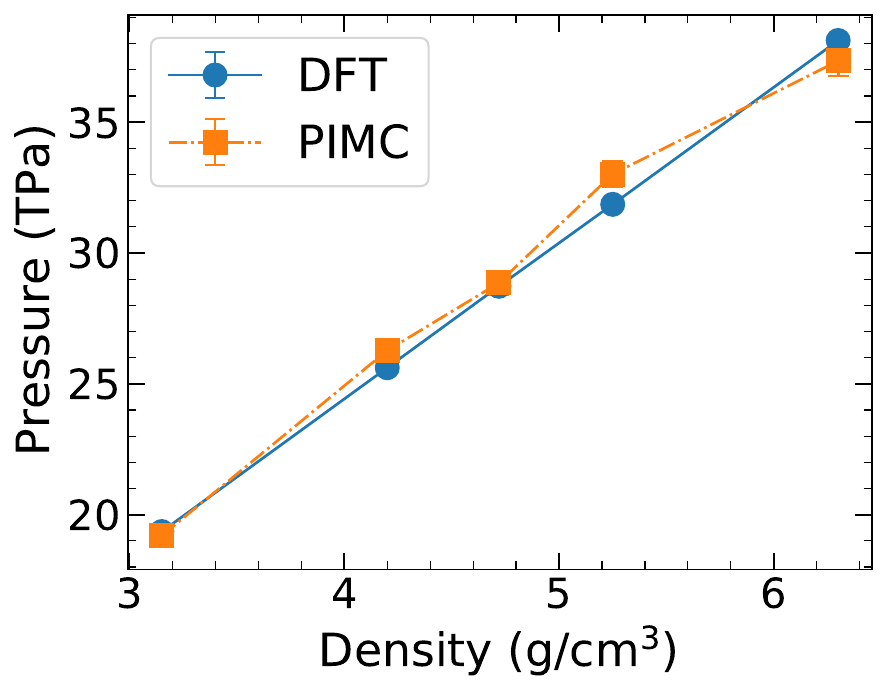}}\hspace{0.01\textwidth}
\subfloat[B\textsubscript{4}C]{\includegraphics[keepaspectratio=true,width=0.32\textwidth]{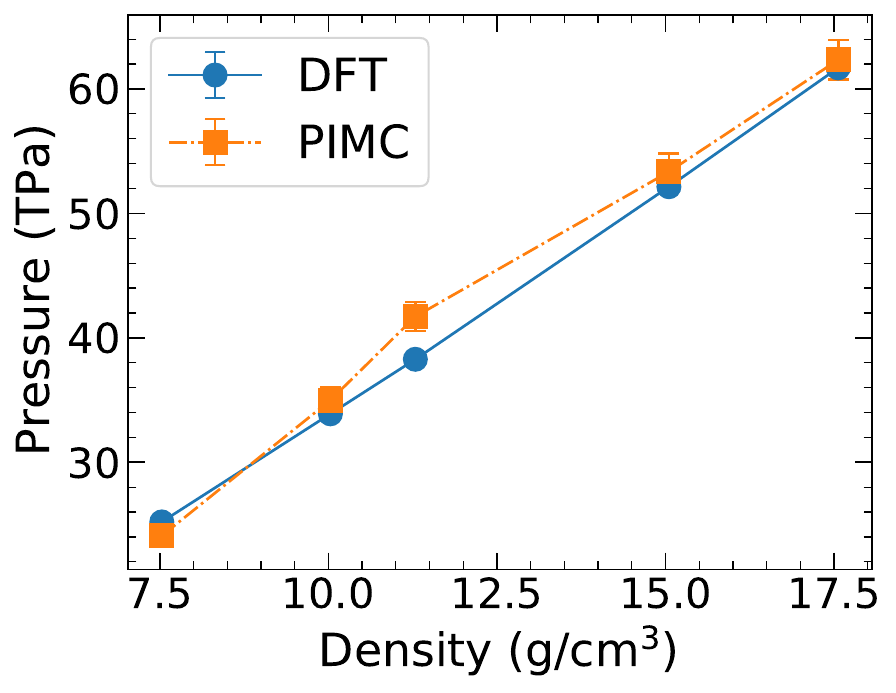}} \hspace{0.01\textwidth}
 \subfloat[MgSiO\textsubscript{3}]{\includegraphics[keepaspectratio=true,width=0.32\textwidth]{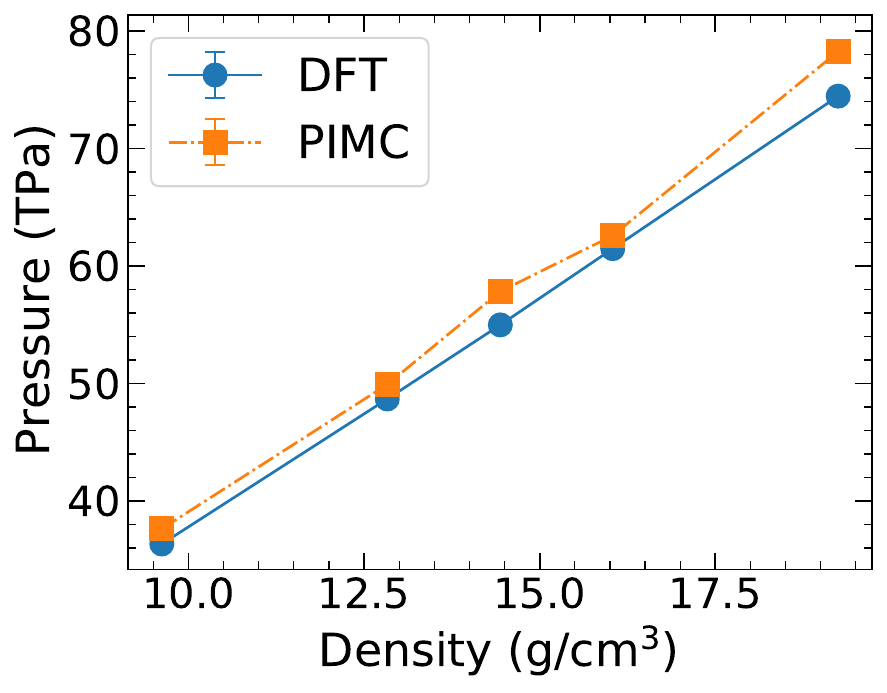}}

 \caption{Variation of pressure with density as computed by DFT (LDA) and PIMC \cite{militzer2021first}.}

 \label{Fig:SmoothVariation}
 \end{figure*}

The results so obtained  are presented in Table~\ref{tab:PIMC_SQ_table},  from which we can make the following observations. First, there is excellent agreement between the DFT results obtained and those in recent work for BN \cite{zhang2019equation}, confirming the accuracy of the simulations. Second, there is very good agreement between DFT and PIMC in the pressure, with a root mean square difference of $\sim 2.5$\%, the larger  differences occurring for systems with larger PIMC error bars.  This level of agreement is notable given the fundamental differences between the two first principles methods; e.g., while PIMC calculations typically employ the fixed node approximation, they are free of the exchange-correlation, self-interaction, and pseudopotential approximations in the DFT calculations. The internal energy computed in DFT and PIMC is also found to be in good agreement. Third, the error bars in DFT are noticeably tighter than PIMC, with root mean square pressure error bar of 0.04\% and 1.5\% for DFT and PIMC, respectively. This translates to significantly smoother variation in the DFT results, as evident from Fig.~\ref{Fig:SmoothVariation}, where the pressure has been plotted as a function of density for B, C, BN, CH, B$_{4}$C, and MgSiO$_{3}$. In particular, the overall trend in the DFT pressure is in agreement with that of the PIMC pressure, which is otherwise more jagged. Overall, the results demonstrate excellent agreement between well-converged DFT (LDA) and PIMC ab initio results for a wide range of WDM/HDM systems --- the average ionization of ions, Coulomb coupling parameter, and electron degeneracy parameter values for these systems can be found in Appendix~\ref{App:Plasma parameters} --- despite the different approximations employed in each.

\begin{table*}[htbp!]
\caption{\label{tab:PIMC_SQ_table} Pressure and internal energy as computed by DFT (LDA) and PIMC \cite{militzer2021first}. The internal energy for each material system is shifted by the mean across the different densities.}
\begin{ruledtabular}
\begin{tabular}{cccccccc}
& & & & \multicolumn{2}{c}{Pressure (GPa)} & \multicolumn{2}{c}{Internal energy (ha/fu)}\\
\cline{5-6} \cline{7-8}
 \multirow{2}{*}{Material} & {Density} & {Comp.} & {Temperature} &  \multirow{2}{*}{DFT} &  \multirow{2}{*}{PIMC} &  \multirow{2}{*}{DFT} &  \multirow{2}{*}{PIMC} \\ 
  & (g/cm\textsuperscript{3}) & {ratio} & (K) & &  & & \\
\hline

 H  & 0.25  & 2.961 & 1 000 000  & 4 082 $\pm$ 1.8  & 4 094 $\pm$ 3 & 0.071 $\pm$ 0.008 &0.078 $\pm$ 0.007\\
 

H & 0.42 & 4.901 & 1 000 000 & 6 730 $\pm$ 1.1& 6 744 $\pm$ 6  & -0.016 $\pm$ 0.003	    &  -0.017 $\pm$ 0.008\\


H    &  0.50 & 5.884 &1 000 000  & 8 068 $\pm$ 1.8  &  8 080 $\pm$ 7& -0.055 $\pm$ 0.004 	    &  -0.061 $\pm$ 0.009\\


He      &        0.39  & 3.135 & 1 000 000  & 2 297 $\pm$ 0.8 &   2 308 $\pm$ 1 & 0.166 $\pm$ 0.009   	    &   0.173 $\pm$ 0.004\\

He      &        0.50  & 4.071 &  1 000 000  & 2 965 $\pm$ 0.9 &   2 981 $\pm$ 1 & 0.003 $\pm$ 0.009  	    &   0.013 $\pm$ 0.003\\

He      &        0.67  & 5.418 &  1 000 000  & 3 930 $\pm$ 0.9 &   3 941 $\pm$ 1  & -0.169 $\pm$ 0.005 	    &   -0.186 $\pm$ 0.003\\

B       &        7.40  & 3.000 & 1 010 479  & 24 839 $\pm$ 8 &  24 394 $\pm$ 259 & 1.065 $\pm$ 0.005   	&     1.089 $\pm$ 0.217\\

B       &        9.86  & 4.000 & 1 010 479  & 33 306 $\pm$ 13 &  32 938 $\pm$ 370& 0.172 $\pm$ 0.005   	&     0.357 $\pm$ 0.232\\

B       &       11.09  & 4.500 &  1 010 479  & 37 671 $\pm$ 18&  36 229 $\pm$ 419  & -0.115 $\pm$ 0.007 	&    -0.496 $\pm$ 0.233\\

B       &       12.33  & 5.000 & 1 010 479  & 42 131 $\pm$ 20 &  41 519 $\pm$ 466  & -0.359 $\pm$ 0.007 	&     -0.250 $\pm$ 0.234\\

B       &       14.79  & 6.000 & 1 010 479  & 51 069 $\pm$ 25&  50 185 $\pm$ 602 & -0.763 $\pm$ 0.007  	&     -0.700 $\pm$ 0.251\\

C       &       8.50  & 2.415 &  1 010 479  & 28 266 $\pm$ 12&  28 519 $\pm$ 32& 0.836 $\pm$ 0.007  	&    0.908 $\pm$ 0.016\\

C       &       10.33  & 2.938 &  1 010 479  & 34 666 $\pm$ 16  &  35 049 $\pm$ 40& 0.325 $\pm$ 0.009 	&    0.369 $\pm$ 0.018\\

C       &       12.64  & 3.590 &  1 010 479  & 42 942 $\pm$ 20  &  43 579 $\pm$ 52& -0.110 $\pm$ 0.009 	&    -0.088 $\pm$ 0.018\\

C       &       15.50  & 4.407 & 1 010 479  & 53 739 $\pm$ 24 &   54 665 $\pm$ 57 & -0.413 $\pm$ 0.010 	&    -0.456 $\pm$ 0.016\\

C       &       19.37  & 5.509 & 1 010 479  & 68 910 $\pm$ 34&  70 512 $\pm$ 102 & -0.638 $\pm$ 0.011  	&    -0.733 $\pm$ 0.023\\

N       &        2.53  & 3.129 &  998 004  & 7 767 $\pm$ 3 &   7 727 $\pm$ 9 & 0.759 $\pm$ 0.010 	    &   1.631 $\pm$ 0.051\\

N       &        3.71  & 4.588 &  998 004  & 11 333 $\pm$ 4&  11 155 $\pm$ 21 & -0.759 $\pm$ 0.009  	&    -1.631 $\pm$ 0.060\\

O       &        2.49  & 3.727 &  998 004  & 7 328 $\pm$ 3&   7 337 $\pm$ 10 & 0.763 $\pm$ 0.009  	 &   0.747 $\pm$ 0.040\\

O       &        3.63  & 5.442 &  998 004  & 10 626 $\pm$ 5&  10 570 $\pm$ 14 & -0.763 $\pm$ 0.010 	&    -0.747 $\pm$ 0.042\\

Ne      &        3.73  & 2.473  & 998 004  & 9 956 $\pm$ 4 &   9 896 $\pm$ 20 & 2.223 $\pm$ 0.007  	&    2.166 $\pm$ 0.084\\

Ne      &        7.90  &  5.239 & 998 004  & 20 931 $\pm$ 10  &  20 731 $\pm$ 46& -2.223 $\pm$ 0.007 	&    -2.166 $\pm$ 0.073\\

Mg      &        4.31  & 2.478 & 1 010 479  & 10 455 $\pm$ 4  &  10 695 $\pm$ 300&  --- 	&   --- \\

LiF     &        7.58  & 2.877  & 998 004  & 22 955 $\pm$ 10 &  22 573 $\pm$ 25 & 1.757 $\pm$ 0.014  	&    1.127 $\pm$ 0.049\\

LiF     &       15.70  & 5.958 &  998 004  & 49 009 $\pm$ 23&  48 415 $\pm$ 42 & -1.757 $\pm$ 0.018  	&    -1.127 $\pm$ 0.039\\

BN  &   6.77  & 3.000 & 1 010 479  & 21 813 $\pm$ 9&  21 447 $\pm$ 629 & 2.057 $\pm$ 0.014     &  1.482 $\pm$ 1.317\\


BN    &   9.03  & 4.000 & 1 010 479  & 29 295 $\pm$ 13& 28 664 $\pm$ 775 & 0.366 $\pm$ 0.018    &  -0.269 $\pm$ 1.217\\


BN     &   10.16  & 4.500 &  1 010 479  & 33 052 $\pm$ 15 &  32 070 $\pm$ 861& -0.234 $\pm$ 0.012     &  -1.419 $\pm$ 1.201\\


BN     &  11.29  & 5.000 & 1 010 479  & 37 019 $\pm$ 17 &  36 758 $\pm$ 462 & -0.707 $\pm$ 0.020    &  -0.456 $\pm$ 1.201\\


BN    &  13.55 & 6.000 &  1 010 479  & 44 949 $\pm$ 22&  46 719 $\pm$ 1087 & -1.482 $\pm$ 0.020    &  0.664 $\pm$ 1.138\\



CH      &        3.15  & 3.000 & 1 347 305  & 19 355 $\pm$ 8 &  19 198 $\pm$ 326  & 2.541 $\pm$ 0.018	&    1.822 $\pm$ 0.770\\

CH      &        4.20  & 4.000 & 1 347 305  & 25 616 $\pm$ 10 &  26 274 $\pm$ 394& 0.557 $\pm$ 0.016  	&    1.563 $\pm$ 0.700\\

CH      &        4.72  & 4.500 & 1 347 305  & 28 728 $\pm$ 12 &  28 857 $\pm$ 445 & -0.213 $\pm$ 0.014 	&    -0.243 $\pm$ 0.701\\

CH      &        5.25  & 5.000 & 1 347 305  & 31 860 $\pm$ 14 &  33 015 $\pm$ 500  & -0.875 $\pm$ 0.012	&   0.444 $\pm$ 0.710\\

CH      &        6.30  & 6.000 & 1 347 305  & 38 123 $\pm$ 10 &  37 343 $\pm$ 572& -2.010 $\pm$ 0.012  	&   -3.586 $\pm$ 0.675\\

B\textsubscript{4}C     &    7.53 & 3.000 &   1 010 479  & 25 208 $\pm$ 10&  24 093 $\pm$ 676  & 5.686 $\pm$ 0.030 	&    -1.656 $\pm$ 2.836\\

B\textsubscript{4}C     &    10.03 & 4.000 &   1 010 479  & 33 893 $\pm$ 15 &  34 986 $\pm$ 1044 & 1.511 $\pm$ 0.040  	&    2.301 $\pm$ 3.285\\

B\textsubscript{4}C     &    11.29 & 4.500 &   1 010 479  & 38 275 $\pm$ 19  &  41 709 $\pm$ 1143& -0.134 $\pm$ 0.045 	&   7.075 $\pm$ 3.197\\

B\textsubscript{4}C     &  15.05  & 6.000 &  1 010 479  & 52 141 $\pm$ 23 &  53 347 $\pm$ 1466 & -2.944 $\pm$ 0.040  	&    -2.840 $\pm$ 3.077\\

B\textsubscript{4}C     &    17.56 & 7.000 &   1 010 479  & 61 691 $\pm$ 30&  62 336 $\pm$ 1580 & -4.119 $\pm$ 0.045  	&   -4.880 $\pm$ 2.836\\

MgSiO\textsubscript{3}  &  9.62 & 3.000 & 1 347 305  & 36 342 $\pm$ 17&  37 637 $\pm$ 486  & 12.262 $\pm$ 0.060 	&   12.518 $\pm$ 2.892\\

 
MgSiO\textsubscript{3}  &  12.83 & 4.000 & 1 347 305  & 48 687 $\pm$  19&  49 919 $\pm$ 632  & 2.387 $\pm$ 0.065 	&  -0.508 $\pm$ 2.821\\

 
MgSiO\textsubscript{3}  & 14.44  & 4.500 & 1 347 305  & 54 993 $\pm$ 19 &  57 845 $\pm$ 716 &  -1.298 $\pm$ 0.050  	&   2.085 $\pm$ 2.838\\

 
MgSiO\textsubscript{3}  &  16.04 & 5.000 & 1 347 305  & 61 481 $\pm$  22 & 62 594 $\pm$ 792 & -4.153 $\pm$ 0.080 	&  -7.665 $\pm$ 2.830\\
 
MgSiO\textsubscript{3}  &  19.25  & 6.000 & 1 347 305  & 74 466 $\pm$ 35&  78 261 $\pm$ 967 & -9.198 $\pm$ 0.065  	&  -6.429 $\pm$ 2.877\\

\end{tabular}
\end{ruledtabular}
\end{table*}


\begin{table*}[htbp!]
\caption{\label{tab:PIMC_SQ_table_difference} Differences in pressure and internal energy for various exchange-correlation functionals, compared to LDA results. The PBE results are obtained from AIMD simulations, while the r$^2$SCAN and PBE0 results are obtained using the $\Delta$-MLFF scheme.}
\begin{ruledtabular}
\begin{tabular}{cccccccccccc}
& & & & \multicolumn{3}{c}{Pressure (\%)} & \multicolumn{3}{c}{Internal energy (ha/fu)}\\
\cline{5-7} \cline{8-10}
 \multirow{2}{*}{Material} & {Density} & {Comp.} & {Temperature} &  \multirow{2}{*}{PBE} &  \multirow{2}{*}{r$^2$SCAN} &  \multirow{2}{*}{PBE0}& \multirow{2}{*}{PBE} &  \multirow{2}{*}{r$^2$SCAN} & \multirow{2}{*}{PBE0}\\ 
  & (g/cm\textsuperscript{3}) & {ratio} & (K) & &  & & \\
\hline
 
H &  0.25 & 2.961 & 1 000 000  & -0.05  & 0.39 & 0.61 & -0.004 & -0.008 & 0.02\\
 
H & 0.42  & 4.901 & 1 000 000 & 0.00 & 0.13 & 0.48 & 0.001 & 0.003 & 0.006 \\

H & 0.50  & 5.884 & 1 000 000 & 0.01 & 0.17 & 0.64 & 0.003 & 0.006 & 0.02 \\
BN & 6.77  & 3.000 & 1 010 479 & -0.05  & -0.23 & -0.41 & -0.050 & -0.062 & -0.072 \\

BN & 9.03  & 4.000 & 1 010 479  & 0.23  & 0.34 & 0.44 & 0.029 & 0.033 & 0.046 \\

BN & 10.16  & 4.500 & 1 010 479  & 0.40  & 0.46 & 0.74 & 0.016 & 0.026 & 0.055 \\

BN & 11.29  & 5.000 & 1 010 479  & 0.08  & 0.39 & 0.62 & -0.037 & -0.047 & -0.052 \\

BN & 13.55  & 6.000 & 1 010 479  & 0.39 & 0.53 & 0.81 & 0.042 & 0.044 & 0.057\\
MgSiO\textsubscript{3} & 9.62  & 3.000 & 1 347 305  & 0.22 & 0.29 & 0.61 & -0.009 & -0.014 & -0.031 \\

MgSiO\textsubscript{3} & 9.62  & 3.000 & 1 347 305  & 0.17 & 0.37 & 0.73 & -0.042 & -0.058 & -0.071 \\

MgSiO\textsubscript{3} & 9.62  & 3.000 & 1 347 305  & 0.24 & 0.24 & 0.81 & 0.122 & 0.145 & 0.214 \\

MgSiO\textsubscript{3} & 9.62  & 3.000 & 1 347 305  & 0.01 & 0.18 & 0.57 & -0.219 & -0.287 & -0.309 \\

MgSiO\textsubscript{3} & 9.62  & 3.000 & 1 347 305  & 0.22 & 0.46 & 0.75 & 0.148 & 0.154 & 0.211 \\

\end{tabular}
\end{ruledtabular}
\end{table*}


\begin{table*}[htbp!]
\caption{\label{tab:Z_gamma_theta_table} Average ionization of ions ($Z^*$), Coulomb coupling parameter ($\Gamma$), and electron degeneracy parameter ($\Theta$), as computed by DFT (LDA).}
\begin{ruledtabular}
\begin{tabular}{cccccccc}
 \multirow{2}{*}{Material} & Density& Comp. & Temperature &  \multirow{2}{*}{$Z^*$} &  \multirow{2}{*}{$\Gamma$} &  \multirow{2}{*}{$\Theta$}  \\ 
 & (g/cm\textsuperscript{3}) & ratio & ($K$) & & \\
\hline
 
 H  & 0.25  & 2.961 & 1 000 000  &  0.96 & 0.13 & 8.68\\

H & 0.42 & 4.901 & 1 000 000 & 0.95 & 0.15 & 6.22\\

H    &  0.50 & 5.884 &1 000 000  & 0.94 & 0.16 & 5.57 \\

He      &        0.39  & 3.135 & 1 000 000  & 1.83 & 0.35 & 10.55  \\

He      &        0.50  & 4.071 &  1 000 000  & 1.86 & 0.39 & 8.86    \\

He      &        0.67  & 5.418 &  1 000 000  & 1.86 & 0.43 & 7.28  \\

B       &        7.40  & 3.000 & 1 010 479  & 2.97 & 1.75 & 2.10  \\

B       &        9.86  & 4.000 & 1 010 479  & 2.78 &  1.69 &  1.82 \\

B       &       11.09  & 4.500 &  1 010 479  & 2.69 & 1.65 & 1.71  \\

B       &       12.33  & 5.000 & 1 010 479  & 2.62 &  1.62 & 1.63   \\

B       &       14.79  & 6.000 & 1 010 479  & 2.49 &  1.55 &  1.49 \\

C       &       8.50  & 2.415 &  1 010 479  & 3.25 &  2.12 & 1.94  \\

C       &       10.33  & 2.938 &  1 010 479  & 3.12 & 2.09 & 1.75  \\

C       &       12.64  & 3.590 &  1 010 479  & 2.94 & 1.99 & 1.59  \\

C       &       15.50  & 4.407 & 1 010 479  & 2.82 & 1.94 & 1.43  \\

C       &       19.37  & 5.509 & 1 010 479  & 2.76 & 1.92 & 1.36  \\

N       &        2.53  & 3.129 &  998 004  & 4.37 & 2.47   & 3.90  \\

N       &        3.71  & 4.588 &  998 004  & 4.29 & 2.69  & 3.06  \\

O       &        2.49  & 3.727 &  998 004  & 4.88 & 2.92 & 4.01  \\

O       &        3.63  & 5.442 &  998 004  & 4.74 & 3.13  &  3.18 \\

Ne      &        3.73  & 2.473  & 998 004  & 5.70 & 4.22 &  3.22 \\

Ne      &        7.90  &  5.239 & 998 004  & 5.19& 4.50 &  2.08 \\

Mg      &        4.31  & 2.478 & 1 010 479  & 6.25 & 4.94 &  3.15 \\

LiF     &        7.58  & 2.877  & 998 004  & 3.48 & 2.31 &  2.08 \\

LiF     &       15.70  & 5.958 &  998 004  & 2.95 & 2.12 & 1.43   \\

BN  &   6.77  & 3.000 & 1 010 479  & 3.46 & 2.21 &  2.22 \\

BN    &   9.03  & 4.000 & 1 010 479  & 3.25 & 2.14 &  1.90 \\

BN     &   10.16  & 4.500 &  1 010 479  & 3.17 & 2.11 & 1.79  \\

BN     &  11.29  & 5.000 & 1 010 479  & 3.09 & 2.09 &  1.69   \\

BN    &  13.55 & 6.000 &  1 010 479  &  2.91 & 1.97 & 1.56\\


CH      &        3.15  & 3.000 & 1 347 305  & 2.62 & 0.87 & 4.24  \\

CH      &        4.20  & 4.000 & 1 347 305  & 2.51 & 0.87 & 3.61  \\

CH      &        4.72  & 4.500 & 1 347 305  & 2.46 & 0.88 & 3.38  \\

CH      &        5.25  & 5.000 & 1 347 305  & 2.42 & 0.88 &  3.18 \\

CH      &        6.30  & 6.000 & 1 347 305  & 2.35 & 0.88 & 2.87  \\

B\textsubscript{4}C     &    7.53 & 3.000 &   1 010 479  & 3.01 & 1.79 & 2.09  \\

B\textsubscript{4}C     &    10.03 & 4.000 &   1 010 479  & 2.84 & 1.77 & 1.79  \\

B\textsubscript{4}C     &    11.29 & 4.500 &   1 010 479  & 2.75 & 1.71 &  1.70 \\

B\textsubscript{4}C     &  15.05  & 6.000 &  1 010 479  & 2.54 & 1.60 &  1.48 \\

B\textsubscript{4}C     &    17.56 & 7.000 &   1 010 479  & 2.42 & 1.54 &  1.38 \\

MgSiO\textsubscript{3}  &  9.62 & 3.000 & 1 347 305  & 5.71 & 4.29 & 2.30  \\
 
MgSiO\textsubscript{3}  &  12.83 & 4.000 & 1 347 305  & 5.35 & 4.17 & 1.98   \\
 
MgSiO\textsubscript{3}  & 14.44  & 4.500 & 1 347 305  & 5.26 & 4.18 & 1.86  \\
 
MgSiO\textsubscript{3}  &  16.04 & 5.000 & 1 347 305  & 4.98 & 3.88 & 1.79  \\
 
MgSiO\textsubscript{3}  &  19.25  & 6.000 & 1 347 305  & 4.89 & 3.98 & 1.61  \\

\end{tabular}
\end{ruledtabular}
\end{table*}


To assess the effect of the exchange-correlation functional, we also consider the temperature-independent semilocal Perdew–Burke–Ernzerhof (PBE) \cite{perdew1996generalized}, semilocal  r$^2$SCAN \cite{furness2020accurate}, and nonlocal PBE0 \cite{adamo1999toward} exchange-correlation functionals for three representative systems: H,  BN, and MgSiO$_{3}$, all with pseudopotentials generated using the PBE exchange-correlation functional. While we still  perform AIMD simulations with SQ-DFT  for the PBE functional, we use the $\Delta$-MLFF models for r$^2$SCAN and PBE0 given their significantly higher computational costs, especially the hybrid PBE0 functional, which includes a fraction of the exact exchange. In particular, we employ LDA as the lower level of theory and use representative configurations from its AIMD trajectory for training. Specifically, we construct a descriptor vector for each configuration encountered in the AIMD using the mean pooling method \cite{jha2018elemnet, antunes2022distributed}, and then apply CUR sparsification \cite{mahoney2009cur} to select 15  configurations with the highest CUR scores, which ensures that the pressure differences predicted by the machine-learned models are converged to within 0.2\%.  The training data corresponding to these 15 configurations is generated by performing  r$^2$SCAN and PBE0 based DFT calculations using the diagonalization and SQ features of SPARC, respectively. The hyperparameters are chosen to be the same as previous work \cite{kumar2024shock, kumar2024fly, kumar2023kohn, SHARMA2025105927}. The $\Delta$-MLFF models so trained are used to calculate the internal energy, atomic forces, and pressure for the same trajectory, i.e., same atomic configurations, as the AIMD with LDA exchange-correlation. In so doing, the need for additional DFT calculations is circumvented, providing significant computational savings. However, it is based on the assumption that the MD trajectories for r$^2$SCAN and PBE0 are sufficiently close to that for LDA. We have verified that this is the case through representative tests. For instance, consider H at a density of 0.5 g/cm$^3$, simulated using an 8-atom cell  for computational efficiency. The errors incurred in computing the pressure using the LDA trajectory for the choice of  r$^2$SCAN and PBE0 exchange-correlation functionals are found to be 0.012\% and 0.024\%, respectively. The corresponding errors when employing the strategy in this work, i.e., choosing 15 representative configurations and using the model so trained to calculate the pressure for the LDA trajectory, are 0.039\%, and  0.066\%,  respectively, further verifying the accuracy of the proposed method. 

We present the results so obtained in Table~\ref{tab:PIMC_SQ_table_difference}, from which we can make the following observations. First, there is very good agreement between the LDA and PBE results, with root mean square differences in pressure and internal energy of 0.2\% and  0.085 ha/fu, respectively.  Second, there is very good agreement between the LDA and r$^2$SCAN results, with root mean square differences in pressure and internal energy of 0.32\% and 0.073 ha/fu, respectively. Third, there is also very good agreement between the LDA and PBE0 results, with root mean square differences in pressure and internal energy of 0.63\% and 0.090 ha/fu, respectively. This quantifies the insensitivity of the results to the choice of exchange-correlation functional at these conditions.  This is consistent with previous findings for comparably well converged Kohn-Sham calculations of BN \cite{zhang2019equation} and C \cite{bonitz2020abinitio} at such conditions considering the LDA and PBE exchange-correlation functionals. 



Overall, the results presented here demonstrate the accuracy of DFT for the study of WDM and HDM. For the temperatures considered here, which are at the upper (lower) end of those encountered in WDM (HDM), it is sufficient for accuracies commonly targeted in practice to use standard temperature-independent local/semilocal exchange-correlation functionals constructed to be exact in the HEG limit. Indeed, the influence of the exchange-correlation is expected to become more significant  at the lower temperatures within the WDM range.


\section{Concluding remarks}
In this work, we have demonstrated the accuracy of Kohn-Sham DFT for WDM and HDM. Specifically, using the SQ method,  we have performed accurate AIMD simulations with temperature-independent local/semilocal density functionals  to determine the equations of state for a wide range of systems at compression ratios of 3x--7x and temperatures around 1 MK. We have found excellent agreement with PIMC benchmarks, the DFT results being smoother and having significantly smaller error bars, demonstrating the accuracy of DFT at such conditions. In addition, using a $\Delta$-MLFF scheme, we have verified that the DFT results are insensitive to the choice of exchange-correlation functional, whether it is local, semilocal, or nonlocal.

The current work further highlights the importance of developing highly efficient methods for the study of WDM/HDM with ab initio accuracy, e.g., machine learned schemes \cite{hinz2023development, liu2020structure, mahmoud2022predicting, kumar2023transferable, chen2024combining, tanaka2022development, kumar2024shock, stanek2024review, kumar2024fly}, making it a worthy subject of further research.


\section*{Acknowledgements}
P.S., X.J., and S.K. gratefully acknowledge support from grant DE-NA0004128 funded by the U.S. Department of Energy (DOE),  National Nuclear Security Administration (NNSA). J.E.P gratefully acknowledges support from the U.S. DOE, NNSA: Advanced Simulation and Computing (ASC) Program at Lawrence Livermore National Laboratory (LLNL). This work was performed in part under the auspices of the U.S. DOE  by LLNL under Contract DE-AC52-07NA27344. 


\appendix
\section{\label{App:Plasma parameters} Average ionization of ions, Coulomb coupling parameter, and electron degeneracy parameter}

In Table~\ref{tab:Z_gamma_theta_table}, we present the average ionization of ions ($Z^*$), Coulomb coupling parameter ($\Gamma$), and electron degeneracy parameter ($\Theta$). In particular, $Z^*$  is calculated using the density of states integration method \cite{callow2023improved}, from the DFT data in the LDA trajectory. Thereafter, the Coulomb coupling and electron degeneracy parameters are computed using the relations \cite{graziani2014frontiers}: 
\begin{align}
\Gamma & = \left(\frac{4\pi n_i}{3}\right)^{1/3} \frac{{Z^*}^2}{k_B T}\,,  \\
\Theta & =  \left(\frac{2}{(3\pi^2n_e)^{2/3}}\right) k_B T\,, 
\end{align}
where  $n_i$ is the ion number density, and $n_e$ is the electron number density. We observe that the $\Gamma$ values lie in the range of 0.13 to 4.29, while the $\Theta$ values lie in the range of 1.36 to 10.55, indicating that the systems are in the WDM/HDM regimes.


\section*{Data Availability Statement}
The data that support the findings of this study are available within the article and from the corresponding author upon reasonable request.

\section*{Author declarations}
The authors have no conflicts to disclose.

\bibliography{sqdft}
    
\end{document}